\documentclass[twocolumn]{aastex62}
\usepackage{amsmath}
\usepackage{lineno}

\graphicspath{{./}{figures/}}

\submitjournal{ApJ}

\shorttitle{Stellar Evolution in AGN Disks}
\shortauthors{Dittmann \& Cantiello}

\begin{document}
\title{A Semi-Analytical Model for Stellar Evolution in AGN Disks}

\correspondingauthor{Alexander J. Dittmann}
\email{dittmann@ias.edu}

\author[0000-0001-6157-6722]{Alexander J.~Dittmann}
\affil{Department of Astronomy and Joint Space-Science Institute, University of Maryland, College Park, MD 20742-2421, USA}
\affil{Theoretical Division, Los Alamos National Laboratory, Los Alamos, NM 87545, USA}

\author[0000-0002-8171-8596]{Matteo Cantiello}
\affil{Center for Computational Astrophysics, Flatiron Institute, 162 5th Avenue, New York, NY 10010, USA}
\affil{Department of Astrophysical Sciences, Princeton University, Princeton, NJ 08544, USA}

\begin{abstract}
Disks of gas accreting onto supermassive black holes may host numerous stellar-mass objects, formed within the disk or captured from a nuclear star cluster. We present a simplified model of stellar evolution applicable to these dense environments; our model exhibits exquisite agreement with full stellar evolution calculations at a minuscule fraction of the cost. Although the model presented here is limited to stars burning hydrogen in their cores, it is sufficient to determine the evolutionary fate of disk-embedded stars: whether they proceed to later stages of nuclear burning and leave behind a compact remnant, reach a quasi-steady state where mass loss and accretion balance one another, or whether accretion proceeds faster than stellar structure can adjust, causing a runaway. We provide numerous examples, highlighting how various disk parameters, and effects such as gap opening, affect stellar evolution outcomes. We also highlight how our model can accommodate time-varying conditions, such as those experienced by a star on an eccentric orbit, and can couple to N-body integrations. This model will enable more detailed studies of stellar populations and their interaction with accretion disks than have previously been possible. 
\end{abstract}


\section{Introduction} \label{sec:intro}
The accretion disks that fuel active galactic nuclei (AGN) \citep[e.g.][]{1964ApJ...140..796S,2008ARA&A..46..475H} may host numerous stars and stellar mass objects. Stars may form within gravitationally unstable regions of these accretion disk \citep[e.g.,][]{1980SvAL....6..357K,2024arXiv240408046H}, or be captured from nuclear star clusters \citep[e.g.,][]{1991MNRAS.250..505S,2020MNRAS.499.2608F}.  

Stellar populations in AGN disks may strongly affect the structure and observable characteristics of the accretion disks around supermassive black holes (SMBHs), particularly in the outer regions of the disk \citep[e.g.,][]{2003MNRAS.341..501S,2024ApJ...966L...9Z}. These populations may beget both electromagnetic \citep[e.g.,][]{1999ApJ...521..502C,2017MNRAS.470.4112G,2021MNRAS.507..156G,2021ApJ...906L...7P} or gravitational wave \citep[e.g.,][]{2017MNRAS.464..946S,2020MNRAS.498.4088M} transients, and could be responsible for the supersolar metal content typically inferred in quasar broad line regions \citep[e.g.,][]{1993ApJ...409..592A,2018MNRAS.480..345X,2022MNRAS.512.2573T}. Closer to home, an accretion disk may have sculpted chemically and dynamically peculiar massive stellar populations in our own galactic center during the most recent accretion episode onto Sagittarius A$^*$ \citep[e.g.,][]{Levin:2003,Paumard:2006,Do:2009,Habibi:2017}. 

Although numerous studies have considered the influences of embedded objects on accretion disks \citep[e.g.,][]{2005ApJ...630..167T,2022ApJ...929..133J,2024ApJ...966L...9Z,2024ApJ...967...88C}, time-dependent models coupling accretion disks, stellar populations, and stellar evolution have yet to be constructed. Thus far, the cost of stellar evolution simulations has prohibited their self-consistent inclusion in such studies, although some stellar evolution studies have yielded insight into the typical fates of embedded stars as functions of various disk properties that can be applied to simplified models \citep[e.g.,][]{2021ApJ...910...94C,2021ApJ...914..105J,2021ApJ...916...48D,2023ApJ...946...56D}. Our hope is that the model presented here or extensions thereof will enable more realistic modeling of the interplay between stars and accretion disks. 

In the following, we present a simple semi-analytical model of stellar evolution within AGN disks, coupling an Eddington standard model to prescriptions for accretion, mass loss, and related processes. We present the resulting set of differential equations in Section \ref{sec:model}. Section \ref{sec:results} provides examples of stellar evolution under various conditions, illustrating the close agreement between our model and various stellar evolution calculations. We discuss some potential future applications, model limitations, and potential extensions in Section \ref{sec:discussion}, and summarize our work in Section \ref{sec:conclusion}. An implementation of our model is available at \href{https://github.com/ajdittmann/starsam}{https://github.com/ajdittmann/starsam}.

\section{Simplified Stellar Models}\label{sec:model}
Our model of stars in AGN disks consists of two main components: equations approximating how the structure of a massive star depends on its bulk properties (its mass and composition); and equations describing how the bulk properties of the star evolve over time (through accretion, mass loss, etc.). Concerning the former, we draw on previous calculations of the structure of massive stars presented in \citet{1984ApJ...280..825B}, which we review in Section \ref{sec:structure}. Concerning the latter, we draw upon the treatments employed in previous full calculations of stellar evolution in AGN disks, such as \citet{2021ApJ...910...94C,2021ApJ...916...48D}, which we review in Section \ref{sec:bulk}. 

\subsection{Stellar Structure}\label{sec:structure}
The fundamental quantities in our stellar models are the total masses of hydrogen, helium, and metals constituting the star ($M_X$, $M_Y$, and $M_Z$), which are used to calculate the total stellar mass and mass fractions 
\begin{align}
M_*=M_X+M_Y+M_Z,\\
X_*=M_X/M_*,~~ Y_*=M_Y/M_*,~~ Z_*=M_Z/M_*.
\end{align}
We assume that the star is chemically homogeneous. In nonrotating stars, this may not be satisfied until $M_*\gtrsim100 M_\odot$, at which point the stars become almost fully convective \citep[see, for example, Figure 3 of][]{2021ApJ...916...48D}. On the other hand, rotating stars are subject to numerous additional instabilities that lead to chemical homogeneity \citep[e.g.,][]{1987A&A...178..159M,2005A&A...443..643Y}. Since stars in AGN disks are likely to spin up as they accrete \citep[e.g.,][]{2021ApJ...914..105J}, the assumption of chemical homogeneity is likely to be well-justified even at lower masses. Homogeneity makes these stars particularly well-described by simplified models (as noted in \citealt{2021ApJ...910...94C}, the structure of stars in AGN disks tend to be well-described by $n=3$ polytropes). In the following, we describe the Eddington standard model for massive stars \citep{1926ics..book.....E}, drawing heavily from Section 2b of \citet{1984ApJ...280..825B}. 

The average number of nuclei and electrons per baryon within the star, the inverse of the mean molecular weight, is given by 
\begin{equation}
Y_T=\mu^{-1}=\frac{1}{4}(6X_*+Y_*+2).
\end{equation}
The photon entropy ($s_\gamma$) is then related to the ratio of radiation pressure to gas pressure ($\sigma\equiv P_{\rm rad}/P_{\rm gas}$) by
\begin{equation}
\sigma = s_\gamma/(4Y_T)=(\beta^{-1}-1), 
\end{equation}
where $\beta\equiv P_{\rm gas}/(P_{\rm gas}+P_{\rm rad})$. In these terms, the familiar form of Eddington's quartic equation 
\begin{equation}
1-\beta=0.003\beta^4\mu^4(M_*/M_\odot)^2
\end{equation}
can be rewritten as 
\begin{equation}
M_* = 1.141 s_\gamma^2(1+4Y_T/s_\gamma)^{3/2} M_\odot.
\end{equation}
The stellar luminosity $L_*$ is related to $s_\gamma$ through the Eddington ratio, $\Gamma \equiv L_*/L_{\rm Edd}$, where
\begin{align}
\Gamma = (4Y_T + 1)^{-1}\\
L_{\rm Edd} \equiv 1.2 \times 10^{38} Y_e^{-1} (M_*/M_\odot)\,{\rm erg\,s^{-1}},
\end{align}
and where $Y_e = (1+X)/2$ is the number of electrons per baryon. Notably, the Eddington standard model discussed here yields an appropriate mass-luminosity relation $L_*\propto M_*^3$ at $\sim$solar masses and $L_*\propto M_*$ at higher masses, and it is our assumption of uniform composition and efficient mixing that breaks down in low-mass stars. 

In order to employ the mass-loss estimates described in Section \ref{sec:bulk}, it is necessary to compute an estimate of the stellar radius. In the ($n=3$) polytropic approximation, the stellar radius is \citep[e.g., Equation 12 of][]{1984ApJ...280..825B}
\begin{equation}
R_* = 30.4Y_T\sigma^{1/2}(1+\sigma)^{1/2}T_c^{-1} R_\odot,
\end{equation}
where $T_c$ is the central temperature of the star in keV. To estimate the central temperatures, we follow appendix A of \citet{1984ApJ...280..825B}, assuming that energy generation is solely due to hydrogen burning via the CNO cycle. Specifically, Equations A2 and A3 of \citet{1984ApJ...280..825B} provide
\begin{align}
\frac{T_c}{2.67}\! =\!(1 \!- \!0.021 \log{A}\! -\! 0.021 \log{X_N}^{-1}\! +\! 0.053\log{T_c})^{-3}\\
A = s_\gamma\left[1+(\nu_p-3)(1+\varsigma)/6\right]^{3/2}\left[XY_e(1+\sigma^{-1})\right]^{-1},\\
\nu_p = 22.42T_c^{-1/3} + 7/3\\
\varsigma = (2s_\gamma/Y_N+1)^{-1}.
\end{align}
The above expressions follow from a power-law fit to the CNO nuclear burning energy generation rate, where $\nu_p$ is the power-law exponent with respect to temperature in the neighborhood of the central temperature and $X_N$ is the nitrogen mass fraction. The $\varsigma$ term is a small modification to the results of \citet{1964ApJS....9..201F} accounting for variation of photon entropy within the convective core \citep{1984ApJ...280..825B}. We approximate the nitrogen mass fraction by scaling the solar \citep[e.g.,][]{2021A&A...653A.141A} nitrogen, carbon, and oxygen abundances by $Z_*/Z_\odot$ and assuming that they are entirely in the form of nitrogen, which is appropriate given the rate-limiting step of the CNO cycle $^{14}_7N+^1_1H\rightarrow ^{15}_8O+\gamma$.

\subsection{Bulk Evolution}\label{sec:bulk}
The bulk composition of the star evolves due to three processes: accretion ($\dot{M}_+$), mass-loss ($\dot{M}_-$), and fusion. In the following, we have specialized to a few specific forms of each of these, but it is trivial to extend our model to other treatments or approximations.  
\subsubsection{Accretion}\label{sec:accretion}
In the following, we have explored four regimes of stellar accretion: spherically symmetric Bondi accretion, tidally-limited accretion, accretion limited by gap formation in an accretion disk, and accretion onto stars with substantial velocity relative to the disk, each of which determine a base accretion rate, $\dot{M}_0$. For a star with zero luminosity, the accretion rate $\dot{M}_+=\dot{M}_0$. However, as the star approaches the Eddington luminosity, the accretion rate is reduced by some factor $0\leq f \leq 1$, according to 
\begin{equation}
\dot{M}_+ = \dot{M}_0f(L_*+L_S),
\end{equation}
where $L_*$ is the intrinsic stellar luminosity and $L_S\approx\dot{M}_+GM_*/R_*$ is the luminosity generated by shocks as accreted material strikes the surface (see Section 4.5 of \citealt{2021ApJ...910...94C}).\footnote{One notational difference between this work and \citet{2021ApJ...910...94C} is that \citet{2021ApJ...910...94C} defined $L_*$ to include both the intrinsic stellar luminosity and the shock luminosity, while here we define these separately. }$^{,}$\footnote{The assumptions made in \citet{2021ApJ...910...94C} may break down when the accretion flow is sufficiently optically thick \citep[e.g.][]{2024arXiv240812017C}.}

When considering spherically symmetric Bondi accretion, the accretion rate is simply given by \citep{1952MNRAS.112..195B}
\begin{align}
\dot{M}_0 = \pi \rho c_s R_B^2\\
R_B \equiv 2GM_*/c_s^2,
\end{align}
where $\rho$ and $c_s$ are the density and sound speed of the AGN disk around the star, and the Bondi radius ($R_B$) characterizes the distance from the embedded star within which the gravity of the star can overcome the thermal response of the fluid. If modeling the evolution of a star moving with significant velocity relative to the background fluid ($v\gtrsim c_s$), then the Bondi-Hoyle-Lyttleton accretion rate may be more appropriate, leading to \citep{1985MNRAS.217..367S,1939PCPS...35..405H}
\begin{equation}\label{eq:bhl}
\dot{M}_0 = 4\pi \rho \frac{G^2M_*^2}{(c_s^2 + v^2)^{3/2}}.
\end{equation}

Within an accretion disk, in the presence of an SMBH, gas simply being within the Bondi radius is not a sufficient condition for it to accrete onto the star, as that gas may still be gravitationally bound to the SMBH. The Hill radius \citep{hill1878researches} defines the region within which gas is bound to the star rather than the SMBH, and can be defined in terms of the angular velocity of the stellar orbit about the SMBH as
\begin{equation}
R_H = \left(\frac{GM_*}{3\Omega^2}\right)^{1/3}.
\end{equation}
For a given $c_s$ and $\Omega$, $R_B < R_H$ at low stellar masses, and $R_H < R_B$ at higher masses. When we consider ``tidally-limited'' accretion, we simply take whichever of these is smaller as the limiting scale, using \citep[e.g.,][]{2020MNRAS.498.2054R,2021ApJ...916...48D}
\begin{equation}\label{eq:mdotTidal}
\dot{M}_0 = \pi \rho c_s \min{\left(R_B^2,R_H^2\right)}.
\end{equation}
It is also possible to generalize Equation (\ref{eq:bhl}) to account for the stellar Hill radius and the vertical extent of the disk, as in \citet{2017MNRAS.464..946S}, prescribing accretion according to 
\begin{align}\label{eq:smh}
\dot{M}_0 = 4\pi \rho R_{\rm acc}\min{\left(R_{\rm acc},H\right)}\\
R_{\rm acc}=\frac{GM_*}{v_{\rm eff}^2}\\
v_{\rm eff} = \left(c_s^2 + R_H^2\Omega^2 + v^2\right)^{1/2}.
\end{align}

If mass ratio between the star and the SMBH ($q\equiv M_*/M_\bullet$) is sufficiently large (for a given disk scale height and effective viscosity) then the presence of the star can begin to alter the large-scale structure of the accretion disk. We consider a simple scenario, where this interaction decreases the average surface density of the disk by some factor $\xi\equiv\Sigma'/\Sigma_0$, the ratio of the perturbed to unperturbed surface densities, which in turn decreases the accretion rate by a factor of $\sim\xi$ \citep[e.g.,][]{2023MNRAS.525.2806C}. We simply approximate that \citep[e.g.,][]{2014ApJ...782...88F,2015MNRAS.448..994K,2020ApJ...891..108D}
\begin{align}
\xi = \frac{1}{1+0.04K}\\
K\equiv q^2/(h^5\alpha), \label{eq:Kdef}
\end{align}
although stars may open even deeper (more depleted) gaps in very low-viscosity disks \citep[e.g.,][]{2018MNRAS.479.1986G}. Thus, if we consider ``gap-limited'' accretion, we take the base accretion rate to be
\begin{equation} \label{eq:gapAcc}
\dot{M}_0 = \pi \rho c_s \min{\left(R_B^2,R_H^2\right)}\xi(\alpha,M_\bullet,h).
\end{equation}

Thus, after choosing some approximation for the unmodified (by radiation) accretion rate, we are left to make some approximation for how radiative feedback further limits accretion. Here, we limit ourselves to an approximation introduced in \citet{2021ApJ...910...94C}, inspired by multidimensional flows where accretion and outflows may coexist. In this case, 
\begin{equation}
f(L_*+L_S) = 1-\tanh{\left[(L_*+L_S)/L_{\rm Edd}\right]}.
\end{equation}
Thus, the accretion rate onto the star can be calculated given its mass, luminosity, and some number of extrinsic parameters dependent on calculation one aims to conduct. 

\subsubsection{Mass Loss}
As stars approach the Eddington limit ($L_*\sim L_{\rm Edd}$), they are thought to lose appreciable mass via continuum-driven winds \citep[e.g.,][]{2004ApJ...616..525O,2014ARA&A..52..487S,2024arXiv240512274C} as radiation pressure overcomes the pull of gravity. In this work, we follow a common approach used in stellar evolution simulations \citep[e.g.,][]{2021ApJ...910...94C,2011ApJS..192....3P}, assuming an outflow at the escape velocity ($v_{\rm esc}=(2GM_*/R_*)^{1/2}$) only when the stellar luminosity is very near the Eddington limit, specifically
\begin{equation}
\dot{M}_- = \frac{L}{v_{\rm esc}}\left[1 + \tanh{\left(\frac{L_*+L_S-L_{\rm Edd}}{0.1L_{\rm Edd}}\right)}\right].
\end{equation}
Although we have only investigated the above prescription in this work, our formalism can accommodate any mass loss prescription that only depends on bulk properties of the star. 

\subsubsection{Nuclear Burning}
We treat nuclear burning in a greatly simplified manner, approximating energy generation due to hydrogen fusion alone. Unlike full stellar evolution calculations, we cannot extend our simplified model beyond core hydrogen burning. A sufficient approximation is that for each helium atom that is produced, four protons are consumed and 27 MeVs are liberated, contributing to the intrinsic stellar luminosity. Thus, helium and hydrogen are created and depleted at a rate
\begin{equation}
\dot{M}_{He,L}=-\dot{M}_{H,L}=\frac{L_*}{27{\rm MeV}/4 m_p},
\end{equation}
where $m_p$ is the proton mass.

\subsection{Piecing the Model Together}
To model the evolution of a star within an AGN disk or other dense environment, one must select initial properties of the star, specifically its initial mass, hydrogen abundance, helium abundance, and metal abundance. One must also select at least an ambient density $\rho$, an ambient sound speed $c_s$, and the mass fractions of the ambient gas ($X, Y, Z$) --- in principle these could be treated as constants, or functions of time. One must then decide which approximation to make of the stellar environment. The above parameters are sufficient to model spherically symmetric  accretion, but if one would like to study stars within accretion disks, one must also chose some angular velocity for the stellar orbit $\Omega$; and possibly a disk aspect ratio $h$, viscosity parameter $\alpha$, and SMBH mass $M_\bullet$ if one aims to account for gap opening --- in our models, we allow $\Omega$ and $h$ to be constants or functions of time, but assume $\alpha$ and $M_\bullet$ to be constant.

Once these choices have been made, simulating stellar evolution in AGN disks boils down to solving a trio of ordinary differential equations (ODEs) describing how the hydrogen, helium, and metal content of the star evolve over time, and at each step solving a series of nonlinear equations to determine the rate of change of each component. Ultimately, we arrive at the system
\begin{align}
\frac{dM_H}{dt} = X\dot{M}_+ - X_*\dot{M}_- - \frac{L_*}{27{\rm MeV}/4 m_p}\\
\frac{dM_{He}}{dt} = Y\dot{M}_+ - Y_*\dot{M}_- + \frac{L_*}{27{\rm MeV}/4 m_p}\\
\frac{dM_Z}{dt} = Z\dot{M}_+ - Z_*\dot{M}_-,
\end{align}
where $L_*$, $\dot{M}_-$, and $\dot{M}_+$ can be calculated following section \ref{sec:structure} and \ref{sec:bulk}. These equations can be solved using standard methods for solving ODEs and a nonlinear one-dimensional root-finding algorithm. 

\begin{figure}
\includegraphics[width=\linewidth]{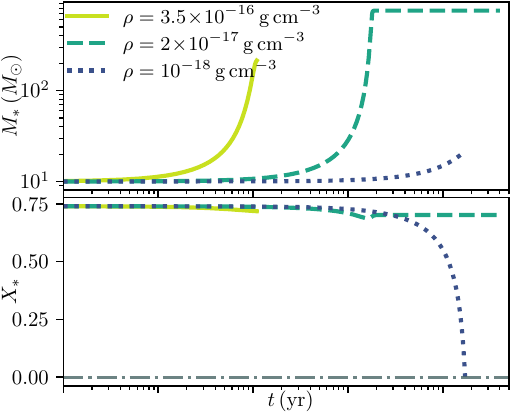}
\caption{Three stellar evolutionary tracks using the model presented in Section \ref{sec:model}, assuming spherically symmetric Bondi accretion and an ambient sound speed of $c_s=10^6\,{\rm cm\,s^{-1}}$. The top panel plots the stellar mass over time and the bottom panel plots the evolution of the stellar hydrogen mass fraction. The stars each began with $M_*=10M_\odot$ and the calculations were differentiated only by the ambient tensity. The lowest-density model, shown by a blue dotted line, ran out of hydrogen and will likely leave behind a compact object. The medium-density model reached a quasi-steady state where mass loss and accretion balance each other. The highest-density model, shown by a solid yellow-green line, underwent runaway accretion. }
\label{fig:outcomes}
\end{figure}

\begin{figure*}
\includegraphics[width=\linewidth]{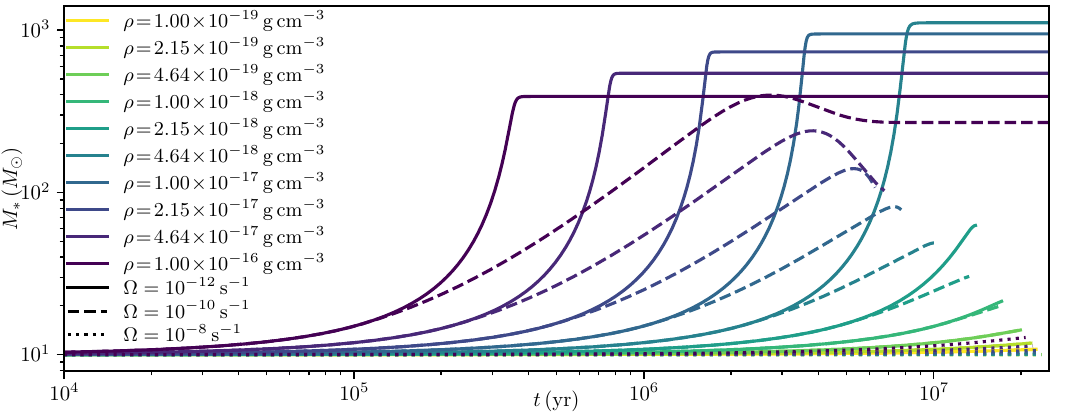}
\caption{Stellar evolution tracks for a range of densities and angular velocities assuming tidally limited accretion as described by Equation (\ref{eq:mdotTidal}). Each model assumed an ambient sound speed of $c_s=10^6\, {\rm cm\, s^{-1}}$. These tracks very nearly approximate full stellar evolution calculations --- see the third row of the second column of Figure 5 in \citet{2021ApJ...916...48D} for a nearly-direct comparison. }
\label{fig:tidal}
\end{figure*}

In addition to a set of ODEs and their initial conditions, we must decide when to terminate each integration. We consider three potential outcomes of stellar evolution. First, stars may simply run out of hydrogen fuel, at which point they would go on to later stages of nuclear burning on a comparatively rapid timescale, likely leaving behind some form of compact remnant \citep[e.g.,][]{2021ApJ...910...94C}; however, the models presented here are not appropriate to stages of stellar evolution beyond core hydrogen burning, so we terminate each integration if $M_H\leq0$. Additionally, it is possible that stellar accretion and mass loss might balance each other, leading to a quasi-steady ``immortal'' state \citep[e.g.][]{2021ApJ...916...48D}; the integration of these models typically ends upon reaching a user-imposed time limit. Additionally, it is possible that stars might accrete more quickly than they can thermally adjust; we do not attempt to model the evolution of these stars subsequent to reaching such a condition, although it seems reasonable that these stars might continue to increase their mass without appreciable radiative feedback until they accrete the entirety of the local gas supply \citep[e.g.,][]{2004ApJ...608..108G}.
In the following calculations, we have estimated the thermal relaxation timescale of massive stars to be 
\begin{equation}
\tau_{\rm KH}=\frac{3}{2}\frac{GM_*^2}{R_*L_*},
\end{equation}
appropriate for a star with $n=3$ polytropic structure. We assume that the star undergoes runaway accretion when $\tau_{\rm acc}\equiv{M_*/\dot{M}_+}<\tau_{\rm KH}$. However, it would be trivial to incorporate a different runaway condition if desired. 
The aforementioned stellar evolution outcomes are illustrated in Figure \ref{fig:outcomes}, where the higher-density model shown in yellow-green undergoes runaway accretion, the medium-density model shown in blue reaches an immortal state, and the lower-density model shown in blue runs out of hydrogen fuel. 

As a word of caution, in some situations it may be unreasonable to terminate calculations when $\tau_{\rm acc}<\tau_{\rm KH}$ without further consideration. For example, for near-solar masses $L_*\propto M^3$ so that $\tau_{\rm KH}$ decreases with mass, but if accretion is limited by tidal effects or gap opening, $\tau_{\rm acc}$ increases with mass. This interplay at low masses might lead to the early termination of a calculation of a star than undergoes a brief period of rapid accretion prior to the accretion timescale increasing. If accretion is limited by both gap opening and tidal effects, the accretion rate ($\dot{M}_0$) itself decreases with stellar mass. 

\section{Examples}\label{sec:results}
We present here a number of example calculations that can be compared to existing stellar evolution calculations. Each of these was computed using a publicly available implementation of our model, which can be found at \href{https://github.com/ajdittmann/starsam}{https://github.com/ajdittmann/starsam}.\footnote{Scripts to reproduce each of our calculations in this paper will be made public upon publication.} 
Unless otherwise noted, we use 
$M_*(t=0)=10M_\odot$, $X_*(t=0)=0.74$, $Y_*(t=0)=0.24$, $X=0.72$, $Y=0.27$, and $c_s=10^6\,{\rm cm\,s^{-1}}$.
In most of the following, we have used the basic RK4(5) integrator \citep{Fehlberg1969} as implemented in scipy \citep{2020SciPy-NMeth}; in Section \ref{sec:2body}, when coupling the above differential equations to a short term orbital integration, we use the standard 4th-order method due to \citet{Kutta}. 

\subsection{Tidally Limited Accretion}\label{sec:tidal}
We begin by comparing our models to existing stellar evolution calculations. In particular, \citet{2021ApJ...916...48D} investigated a number of approximations for how deviation from spherical symmetry in AGN disks could affect stellar evolution, and found that the tidal limiting described by Equation (\ref{eq:mdotTidal}) was often significant and the most widely applicable. 

We present in Figure \ref{fig:tidal} a number of $M_*(t)$ tracks, paralleling the calculations in \citet{2021ApJ...916...48D}, although assuming very slightly different mass fractions. We find very good agreement between the results of our simplified model and full stellar evolution calculations. For example, at both $\Omega=10^{-10}\,{\rm s^{-1}}$ and $\Omega=10^{-12}\,{\rm s^{-1}}$ we find that models transition from exhausting their helium supply to reaching a quasi-steady state at the same ambient density, although the exact shape of the $M_*(t)$ curve deviates slightly in the $\rho=10^{-16}\,{\rm g\, cm^{-3}},\,\Omega=10^{-10}\,{\rm s^{-1}}$ case.

Overall, given the enormous simplifications of the present model, the level of agreement they achieve with full stellar evolution calculations is extremely encouraging. They seem to reproduce the typical evolutionary outcomes of a given star, as well as their masses, to good precision. For example, the masses of immortal stars in Figure \ref{fig:tidal} usually deviate from those in the relevant models of \citet{2021ApJ...916...48D} by factors of at most a factor of a few. 

One perhaps counter-intuitive result displayed in Figure \ref{fig:tidal}, particularly the simulations at $\Omega=10^{-12}\,\rm{s^{-1}}$, is that immortal stars embedded in high-density media reach lower masses than stars embedded in comparatively lower-density media. In these cases, the accretion shock luminosity can reach a substantial fraction of the total stellar luminosity, resulting in substantially higher mass loss rates at a given mass. This trend has been observed in previous full in stellar evolution simulations \citep[e.g.,][]{2021ApJ...910...94C,2021ApJ...916...48D}. 

\subsection{Gap Opening}
The possibility of stars opening gaps in AGN disks and subsequently decreasing the rate of accretion onto embedded stars was discussed in \citet{2021ApJ...916...48D}, although they did not carry out any calculations since Equation (\ref{eq:Kdef}) illustrates that gap opening will only be relevant in disks that are either very thin $h\lesssim 10^{-3}$, nearly inviscid $\alpha \lesssim 10^{-4}$, or those orbiting very low-mass SMBHs $M_\bullet \lesssim 10^6 M_\odot$. Although gap opening is more important for intermediate-mass black holes and extremely massive ($M_*\gtrsim10^4$) stars \citep[e.g.,][]{2004ApJ...608..108G,2012MNRAS.425..460M}, it can also be relevant for stars and stellar-mass black holes around less-massive SMBHs \citep[e.g.,][]{2020MNRAS.493.3732D}.

\begin{figure}
\includegraphics[width=\linewidth]{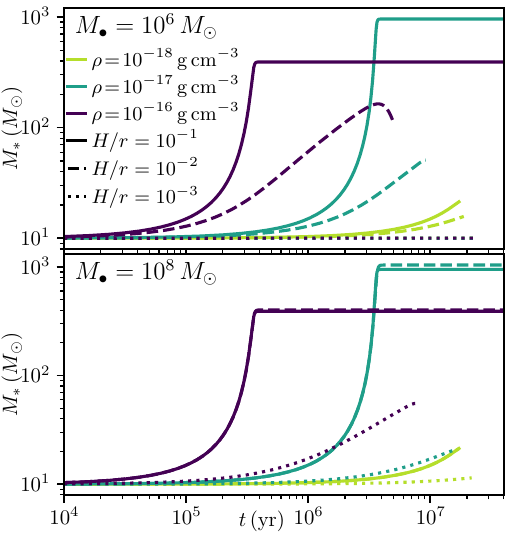}
\caption{Stellar evolution tracks at various disk densities, highlighting the importance of gap opening, as described by Equation \ref{eq:gapAcc}. Each of these calculations fixed $c_s=10^6\,{\rm cm\,s^{-1}},$ $\alpha=0.1$, and $\Omega=10^{-10}\,{\rm s^{-1}}.$ Around lower-mass SMBHs, gap opening can greatly suppress accretion onto embedded stars, while around higher-mass SMBHs gap opening is only relevant in extremely thin disks.}
\label{fig:gaps}
\end{figure}

Figure \ref{fig:gaps} illustrates the effects of gap opening explicitly for disks with $\alpha=0.1,\,c_s=10^6\, {\rm cm\,s^{-1}},$ and $\Omega = 10^{-10}\, {\rm s^{-1}}$. In the top panel, showing calculations assuming an SMBH mass of $10^6\,M_\odot$, the disk aspect ration $h=H/r$ strongly affects the evolution of embedded stars. In the $h=10^{-3}$ case, accretion is entirely inhibited, since initially $K=10^6$. In the $h=10^{-2}$ case, $K$ was initially $\sim10$, so accretion was not suppressed until the stars reached somewhat higher masses. One other hand, in $h=0.1$ disks, stellar evolution was unaffected. 

The bottom panel of Figure \ref{fig:gaps} illustrates the same calculations for a more massive SMBH. In this case, the $h=0.1$ and $h=10^{-2}$ disks lead to nearly indistinguishable stellar evolution, while accretion onto stars in the $h=10^{-3}$ case is heavily suppressed. Intriguingly, in the $\rho=10^{-17}\,{ \rm g\,cm^{-3}}$ case, gap opening actually leads to a higher-mass immortal star by decreasing the accretion rate as discussed in Section \ref{sec:tidal}. It is conceivable that gap opening might lead to disparate stellar populations around low- and high-mass SMBHs. However, we leave a detailed investigation of this and its observational consequences for future studies. 

\begin{figure}
\includegraphics[width=\linewidth]{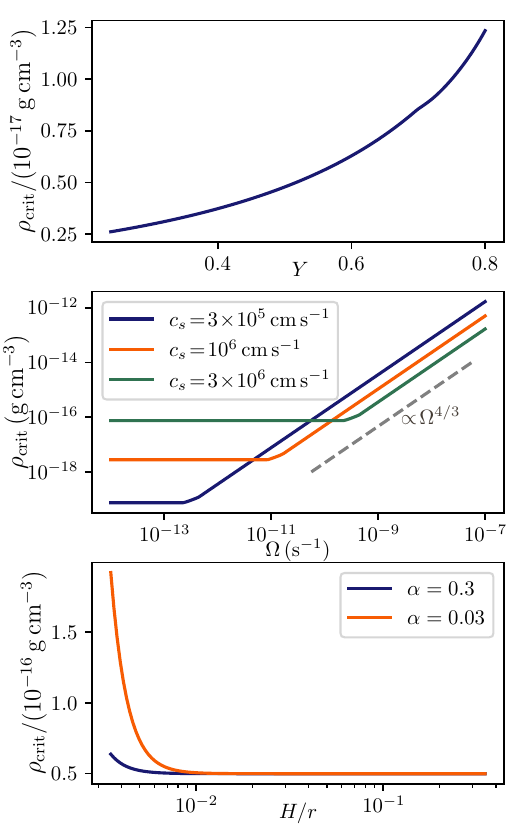}
\caption{The critical density delineating the boundary between those stars reaching quasi-steady states and those that exhaust their fuel supply as a function of various stellar parameters. The top panel assumes spherically symmetric accretion and varies the helium mass fraction of the accreting gas, the middle panel considers tidally limited accretion in disks with various sound speeds, and the bottom panel considers gap opening in disks with varying aspect ratios $h=H/r$ around a $10^8\,M_\odot$ SMBH. Each line is sampled by 120 points, such that this figure incorporates $\gtrsim18000$ calculations.}
\label{fig:rhocrit}
\end{figure}

\subsection{The Critical Density to Sustain Immortal Stars}
One of the questions we hope to answer with our models is what environmental conditions lead to which stellar evolutionary outcomes --- what sort of disk parameters should lead to stars reaching quasi-steady states, which should lead to runaway accretion, and which should should produce stars poised to undergo supernovae? Here, we examine the boundary between stars that reach a quasi-steady state and those that run out of hydrogen fuel. This depends on whether or not the stellar nuclear burning timescale can become shorter than the accretion timescale prior to accretion and mass loss achieving a quasi-steady state (see Figure 1 of \citealt{2023ApJ...946...56D}). Figure \ref{fig:rhocrit} collects calculations determining this critical density in a number of scenarios using 25 bisection iterations at each point.

The top panel of Figure \ref{fig:rhocrit} examines  the effect of the background helium abundance of the AGN disk on the critical density for immortal stars undergoing spherically symmetric (Bondi) accretion. Here, because stars with higher mean molecular weights tend to be more luminous at constant mass than those with lower mean molecular weights, stars that accrete material with a higher helium abundance will have shorter nuclear burning timescales at a given mass, requiring higher densities to achieve a short enough accretion timescale to sustain a star in a quasi-steady state. The results of this calculation agree very closely with the analogous stellar evolution calculations presented as the orange curve in Figure 5 of \citet{2023ApJ...946...56D}. 

The middle panel of Figure \ref{fig:rhocrit} plots the critical density as a function of angular velocity and sound speed for stars undergoing tidally-limited accretion. At sufficiently low angular velocities, accretion occurs in a spherically symmetric manner, requiring much higher densities at high sound speeds, since $R_B\propto c_s^{-2}$. However, at higher angular velocities, the accretion rate is proportional to the sound speed, so lower densities are required to sustain immortal stars at fixed $\Omega$. The orange curve, for $c_s=10^6\,{\rm cm\,s^{-1}}$, agrees very well with the analogous result based on stellar evolution calculations (Figure 8 and Equation 22 of \citealt{2021ApJ...916...48D}) in the $\Omega\gtrsim10^{-11}\,{\rm s^{-1}}$ regime where those simulations were conducted. 

The bottom panel of Figure \ref{fig:rhocrit} shows the critical density as a function of disk aspect ratio and viscosity for a star orbiting a $10^8\,M_\bullet$ SMBH at $\Omega=10^{-10}\,{\rm s^{-1}}$ within a $c_s=10^6\,{\rm cm\,s^{-1}}$ disk. The values of $\alpha$ chosen here roughly bracket those used to model compact object accretion disks \citep[e.g.,][]{2007MNRAS.376.1740K} and the effective values of $\alpha$ derived from global simulations of the magnetorotational instability \citep[e.g.,][]{2016ApJ...826...40H}. At this SMBH mass, gap opening makes it more difficult to sustain stars in a quasi-steady state in very thin disks ($h<10^{-2}$), although it is still quite possible.  

\subsection{A Star on an Eccentric Orbit}\label{sec:2body}
The model presented here should facilitate coupled N-body and stellar evolution calculations. To illustrate this capability, we study here the evolution of a 2-body star-SMBH system and a toy disk model. 
Here, we assume an initially $10\,M_\odot$ star on a prograde eccentric ($e=0.4$) orbit around a $10^7\,M_\odot$ SMBH with a period of $\sim 1\,{\rm Myr}$. The star moves though an isothermal Keplerian accretion disk with $c_s=10^5\,\rm{cm\,s^{-1}}$ and a density profile given by
$\rho(r) = \rho_0(r/r_0)^{-1},$ 
normalized to $3\times10^{-13}\rm{g\,cm^{-3}}$ at 10 AU. 

\begin{figure}
\includegraphics[width=\linewidth]{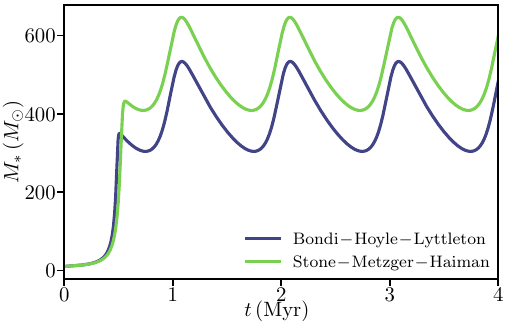}
\caption{The mass of a star orbiting through a toy model of an AGN disk on an eccentric $(e=0.4)$ orbit, comparing two different models of accretion --- results assuming Equation (\ref{eq:bhl}) are shown in dark blue, and results assuming Equation (\ref{eq:smh}) are shown in green. After accreting up to the point of reaching a quasi-steady balance between accretion and mass loss, the stellar mass varies over time due to changes in the disk density and relative velocity between the disk and star. This calculation assumed $X_*=0.7,~Y_*=0.2,$ and $Z_*=0.1$ at $t=0$, and assumed that the background disk had a composition of $X=0.7,~Y=0.26,~Z=0.04$. }
\label{fig:nbody}
\end{figure}

Figure \ref{fig:nbody} illustrates the results of an example calculation over four orbital periods under the assumptions of Equations (\ref{eq:bhl}) and (\ref{eq:smh}), where we have neglected the change in stellar velocity due to accretion or drag forces, which over this dynamically short integration have little effect. In this case, the accretion timescale is shorter than the orbital timescale and the star essentially moves between two quasi-steady states of differing mass. Notably, in a multi-body system, scattering events could easily drive stars into regions where their evolution is irrevocably altered. 

\begin{figure}
\includegraphics[width=\linewidth]{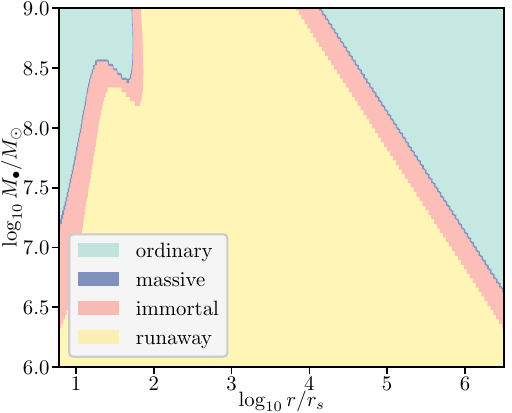}
\caption{The outcomes of a suite of simulations as functions of SMBH mass and the stellar location within the disk (in units of the Schwarzschild radius $r_s\equiv 2GM/c^2$), based on the \citet{2003MNRAS.341..501S} disk models analyzed in \citet{Fabj2024}. This figure illustrates the results of $8\times10^5$ calculations, a task utterly infeasible for full stellar evolution simulations.}
\label{fig:diskmodels}
\end{figure}

\subsection{Disk Models}
Our model will enable more precise feedback between models of AGN disks and the stellar populations they harbor. To illustrate this possibility, albeit without incorporating the effects of stellar feedback on the disk, we illustrate in Figure \ref{fig:diskmodels} the stellar evolutionary outcomes of a set of $8\times10^5$ calculations; each simulation assumed the star to reside at a fixed location in a set of \citet{2003MNRAS.341..501S} disk models calculated in \citet{Fabj2024}, allowed a maximum duration $10^7$ years, and assumed tidally-limited accretion (without considering gap opening). We classify stars as `runaway' ($\tau_{\rm acc}<\tau_{{\rm KH}}$), `immortal' (reaching a quasi-steady state), `massive' (achieving a mass greater than $8 M_\odot$ and accreting at too slow a rate to achieve a quasi-steady state), and `ordinary' otherwise. A shorter calculation, effectively a smaller upper limit on the AGN disk lifetime, would primarily limit the number of higher-mass stars. The calculations in this subsection assumed an initial stellar mass of $M_*=M_\odot$, which greatly increased the initial accretion timescale compared to the calculations in prior sections.

The same general trends occur at each SMBH mass, although the extent of each region varies with mass. Gas densities are typically very low at large radii, usually resulting in very low accretion rates that do not affect stellar evolution significantly. Moving inwards, gas densities increase and the accretion timescale can become comparable or shorter than the initial stellar nuclear burning timescale and stars grow massive. This trend continues, and stars eventually accrete quickly enough to reach quasi-steady states or undergo runaway accretion. However, at smaller radii, the gas density varies non-monotonically, while the disk temperature increases and tidal forces from the SMBH become stronger, limiting accretion. 

Our model predicts slightly fewer massive and immortal stars than the analytical calculations of \citet{Fabj2024} but a similar number of stars undergoing runaway accretion. This is most likely because of the slightly different criteria used to differentiate between `ordinary' and `massive stars,' \citet{Fabj2024} comparing the initial accretion timescale to the expected disk lifetime in contrast to our cut based on the final stellar mass. Notably, more massive stars have shorter accretion timescales, so these results depend on our choice of initial stellar masses. 

\pagebreak
\section{Discussion}\label{sec:discussion}
The models presented in this work, thanks to their close agreement with full stellar evolution simulations at a minuscule fraction of the cost, will enable new avenues of research detailed below, which we hope will facilitate rigorous comparisons between theory and observation. Specifically, these models should make it possible to carry out time-dependent studies of stellar populations in AGN disks with minimal sacrifices compared to full stellar evolution calculations. Furthermore, the model presented here may be extended in a number of ways, to broaden its applicability to other stages of stellar evolution, or to incorporate other physical effects. 
However, when applying the present models, one should keep in mind some of the assumptions used to derive them and the resulting model limitations, which we collect below.

\subsection{Applications}
The model developed here should make it possible to conduct detailed time-dependent studies of stellar populations in AGN disks, circumventing the cost of full stellar evolution simulations. For example, numerous studies have coupled N-body or smoothed-particle-hydrodynamics codes to either full \citep[e.g.,][]{2016MNRAS.455..462V,2016MNRAS.456.3401P} or approximate \citep[e.g.,][]{2021PASJ...73.1074F,2024MNRAS.528.5119A} stellar evolution. Because the model developed here is a relatively straightforward ordinary differential equation, it should be straightforward to couple it to N-body calculations. Presumably, an accretion prescription along the lines of Equation (\ref{eq:bhl}) or Equation (\ref{eq:smh}) would be appropriate for such studies. 

Because the model presented here self-consistently predicts stellar luminosities and mass loss rates given local disk properties, it should improve models of stellar feedback on AGN disks \citep[e.g.][]{2005ApJ...630..167T,2024ApJ...966L...9Z,2024ApJ...967...88C}. It may also be possible to incorporate these stellar models into time-dependent disk model calculations \citep{2022ApJ...928..191G,2024arXiv240509380E}, or one-dimensional Monte Carlo calculations similar to those used to predict stellar-mass black hole merger rates in AGN disks \citep[e.g.][]{2020MNRAS.498.4088M,2020ApJ...898...25T}. 

\subsection{Caveats}
The models presented here rely on numerous assumptions that may cause their predictions to deviate from reality. Many of these, such as the highly approximate prescriptions for how stars accrete and lose mass, and how the accreting matter interacts with a near-Eddington radiation field, follow from the necessity of describing intrinsically three dimensional and turbulent effects within a simplified ODE model. Such ad hoc assumptions are also shared by full stellar evolution calculations; although such assumptions cannot be circumvented, their effects on predictions can be gauged through parameter studies, which are far more efficient using the model presented here. 

Compared to full stellar evolution calculations, the simplified model presented in Section \ref{sec:model} has primarily sacrificed spatial and chemical fidelity. The assumption of CNO-dominated burning limits its applicability to stars more massive than the Sun and with appreciable metallicity (making them ill-suited to studying stars in the very early Universe). The assumption of an $n=3$ polytropic structure is quite suitable for stars embedded in AGN disks \citep{2021ApJ...910...94C}, but is nevertheless a limitation. The assumption of chemical homogeneity is not reasonable for stars of all masses: below $M\lesssim100M_\odot$, nonrotating stars still possess a radiative envelope, which may preclude freshly accreted material from mixing into the stellar core \citep[e.g.,][]{2023MNRAS.526.5824A}. However, stars accreting from AGN disks are likely to reach substantial rotation rates \citep{2021ApJ...914..105J}, at which point rotation-induced mixing should lead to chemical homogeneity at virtually all stellar masses \citep[e.g.][]{1987A&A...178..159M,2005A&A...443..643Y}.

\subsection{Extensions}
Many of the model components presented here, such as the recipes for the accretion rate onto stars given external accretion disk conditions, the interplay between near-Eddington radiation fields and accretion, and continuum-driven mass loss, are quite flexible. Similarly one may choose to implement different criteria for runaway accretion. For example, in the theoretical study of early-Universe supermassive black hole formation, some have considered the balance between the accretion and Kelvin-Helmholtz timescales of the stellar envelope specifically, as opposed to the global timescales employed here \citep[e.g.,][]{2015MNRAS.452..755S}. 

Beyond adjusting the components already present within the model, it is possible to incorporate other considerations entirely. For example, \citet{2024arXiv240812017C} has suggested modifications to the stellar accretion rate applicable when the accretion flow onto the star is very optically thick. Incorporating an accretion rate along the lines suggested in Section 3.2 of \citet{2024arXiv240812017C}, suitably modified when $R_H<R_B$ as discussed in Section \ref{sec:accretion} \citep[and, for example, in][]{2020MNRAS.498.2054R,2023MNRAS.525.2806C}, would be a straightforward way to extend the applicability of our model to other regimes. Judging by Figure \ref{fig:diskmodels}, limiting the stellar accretion rate has the potential to substantially increase the number of massive and immortal stars within the disk

One natural extension of the models presented here is to explicitly account for stellar angular momentum. Thanks to the background shear flow of AGN disks, accretion may spin stars up to considerable angular velocities. Prescriptions for these effects have been incorporated into previous stellar evolution calculations \citep{2021ApJ...914..105J}, which would be straightforward to incorporate into the semi-analytical model presented here by introducing an additional differential equation for the total stellar spin. One crucial ingredient for such considerations is the stellar moment of inertia, which for the Eddington standard model discussed here is approximately 
\begin{equation}
I_*\approx 7.529\times10^{-2} M_*R_*^2.
\end{equation}

One of the limitations of the model presented in this work is its restriction to burning via the CNO cycle. At least under the assumption of chemically homogeneous evolution \citep[e.g.,][]{1987A&A...178..159M,2005A&A...443..643Y} which seems fairly applicable to the evolution of stars in AGN disks \citep[e.g.,][]{2021ApJ...910...94C}, it may be reasonable to assume energy generation due to triple-$\alpha$ burning after hydrogen exhaustion. Such considerations might improve the estimates made by these models of the amount of metals ejected into the ambient medium. However, modeling later stages of stellar evolution would require evolving the abundances of more elements, at which points the costs would likely outweigh potential benefits. 
Moreover, the time a star spends burning heavier elements is significantly shorter than the time it spends burning hydrogen, which usually accounts for 90\% of the stellar lifetime. Later stages of burning make up a smaller fraction of the the lives of AGN stars, where accretion and mixing result in relatively longer hydrogen burning phases.

\section{Conclusions}\label{sec:conclusion}
In this work, we have combined a simple model of stellar structure, an Eddington standard model \citep{1926ics..book.....E,1984ApJ...280..825B}, with approximations for the interplay between stellar radiation, accretion, and mass loss and AGN disks employed in previous stellar evolution calculations \citep{2021ApJ...910...94C,2021ApJ...916...48D} to yield a simple model of stellar evolution in AGN disks or other dense environments. As shown in Section \ref{sec:results}, our model very closely reproduces the results of full stellar evolution simulations, typically at less than a thousandth of the cost, although the reduced model is only applicable to core hydrogen burning via the CNO cycle in its present form. We hope that our distillation of the equations of stellar evolution in AGN disks to a trio of ordinary differential equations will enable future studies to investigate the interplay between stellar populations and their environments in much more detail than had been possible before now. 

\section*{Software}
\texttt{matplotlib} \citep{4160265}, \texttt{numpy} \citep{5725236}, scipy \citep{2020SciPy-NMeth}

\section*{Acknowledgments}
AJD is grateful for the hospitality of the Center for Computational Astrophysics throughout a visit during which this work was initiated.
The Center for Computational Astrophysics at the Flatiron Institute is supported by the Simons Foundation.
AJD gratefully acknowledges support from LANL/LDRD under project number 20220087DR, and NASA ADAP grants 80NSSC21K0649 and 80NSSC20K0288. The LA-UR number is LA-UR-24-28455.

\bibliographystyle{aasjournal}
\bibliography{references}

\begin{thebibliography}{}
\expandafter\ifx\csname natexlab\endcsname\relax\def\natexlab#1{#1}\fi
\providecommand{\url}[1]{\href{#1}{#1}}

\bibitem[{{Ali-Dib} \& {Lin}(2023)}]{2023MNRAS.526.5824A}
{Ali-Dib}, M., \& {Lin}, D. N.~C. 2023, \mnras, 526, 5824

\bibitem[{{Arca Sedda} {et~al.}(2024){Arca Sedda}, {Kamlah}, {Spurzem},
  {Giersz}, {Berczik}, {Rastello}, {Iorio}, {Mapelli}, {Gatto}, \&
  {Grebel}}]{2024MNRAS.528.5119A}
{Arca Sedda}, M., {Kamlah}, A. W.~H., {Spurzem}, R., {et~al.} 2024, \mnras,
  528, 5119

\bibitem[{{Artymowicz} {et~al.}(1993){Artymowicz}, {Lin}, \&
  {Wampler}}]{1993ApJ...409..592A}
{Artymowicz}, P., {Lin}, D.~N.~C., \& {Wampler}, E.~J. 1993, \apj, 409, 592

\bibitem[{{Asplund} {et~al.}(2021){Asplund}, {Amarsi}, \&
  {Grevesse}}]{2021A&A...653A.141A}
{Asplund}, M., {Amarsi}, A.~M., \& {Grevesse}, N. 2021, \aap, 653, A141

\bibitem[{{Bond} {et~al.}(1984){Bond}, {Arnett}, \&
  {Carr}}]{1984ApJ...280..825B}
{Bond}, J.~R., {Arnett}, W.~D., \& {Carr}, B.~J. 1984, \apj, 280, 825

\bibitem[{{Bondi}(1952)}]{1952MNRAS.112..195B}
{Bondi}, H. 1952, \mnras, 112, 195

\bibitem[{{Cantiello} {et~al.}(2021){Cantiello}, {Jermyn}, \&
  {Lin}}]{2021ApJ...910...94C}
{Cantiello}, M., {Jermyn}, A.~S., \& {Lin}, D. N.~C. 2021, \apj, 910, 94

\bibitem[{{Chen} {et~al.}(2024){Chen}, {Jiang}, {Goodman}, \&
  {Lin}}]{2024arXiv240812017C}
{Chen}, Y.-X., {Jiang}, Y.-F., {Goodman}, J., \& {Lin}, D. N.~C. 2024, arXiv
  e-prints, arXiv:2408.12017

\bibitem[{{Chen} \& {Lin}(2024)}]{2024ApJ...967...88C}
{Chen}, Y.-X., \& {Lin}, D. N.~C. 2024, \apj, 967, 88

\bibitem[{{Cheng} \& {Wang}(1999)}]{1999ApJ...521..502C}
{Cheng}, K.~S., \& {Wang}, J.-M. 1999, \apj, 521, 502

\bibitem[{{Cheng} {et~al.}(2024){Cheng}, {Goldberg}, {Cantiello}, {Bauer},
  {Renzo}, \& {Conroy}}]{2024arXiv240512274C}
{Cheng}, S.~J., {Goldberg}, J.~A., {Cantiello}, M., {et~al.} 2024, arXiv
  e-prints, arXiv:2405.12274

\bibitem[{{Choksi} {et~al.}(2023){Choksi}, {Chiang}, {Fung}, \&
  {Zhu}}]{2023MNRAS.525.2806C}
{Choksi}, N., {Chiang}, E., {Fung}, J., \& {Zhu}, Z. 2023, \mnras, 525, 2806

\bibitem[{{Dempsey} {et~al.}(2020){Dempsey}, {Lee}, \&
  {Lithwick}}]{2020ApJ...891..108D}
{Dempsey}, A.~M., {Lee}, W.-K., \& {Lithwick}, Y. 2020, \apj, 891, 108

\bibitem[{{Dittmann} {et~al.}(2021){Dittmann}, {Cantiello}, \&
  {Jermyn}}]{2021ApJ...916...48D}
{Dittmann}, A.~J., {Cantiello}, M., \& {Jermyn}, A.~S. 2021, \apj, 916, 48

\bibitem[{{Dittmann} {et~al.}(2023){Dittmann}, {Jermyn}, \&
  {Cantiello}}]{2023ApJ...946...56D}
{Dittmann}, A.~J., {Jermyn}, A.~S., \& {Cantiello}, M. 2023, \apj, 946, 56

\bibitem[{{Dittmann} \& {Miller}(2020)}]{2020MNRAS.493.3732D}
{Dittmann}, A.~J., \& {Miller}, M.~C. 2020, \mnras, 493, 3732

\bibitem[{{Do} {et~al.}(2009){Do}, {Ghez}, {Morris}, {Lu}, {Matthews}, {Yelda},
  \& {Larkin}}]{Do:2009}
{Do}, T., {Ghez}, A.~M., {Morris}, M.~R., {et~al.} 2009, \apj, 703, 1323

\bibitem[{{Eddington}(1926)}]{1926ics..book.....E}
{Eddington}, A.~S. 1926, {The Internal Constitution of the Stars}

\bibitem[{{Epstein-Martin} {et~al.}(2024){Epstein-Martin}, {Tagawa}, {Haiman},
  \& {Perna}}]{2024arXiv240509380E}
{Epstein-Martin}, M., {Tagawa}, H., {Haiman}, Z., \& {Perna}, R. 2024, arXiv
  e-prints, arXiv:2405.09380

\bibitem[{{Fabj} {et~al.}(2024){Fabj}, {Dittmann}, {Cantiello}, {Perna}, \&
  {Samsing}}]{Fabj2024}
{Fabj}, G., {Dittmann}, A.~J., {Cantiello}, M., {Perna}, R., \& {Samsing}, J.
  2024, arXiv e-prints, arXiv:2408.16050

\bibitem[{{Fabj} {et~al.}(2020){Fabj}, {Nasim}, {Caban}, {Ford}, {McKernan}, \&
  {Bellovary}}]{2020MNRAS.499.2608F}
{Fabj}, G., {Nasim}, S.~S., {Caban}, F., {et~al.} 2020, \mnras, 499, 2608

\bibitem[{{Fehlberg}(1969)}]{Fehlberg1969}
{Fehlberg}, E. 1969, {Low-order classical Runge-Kutta formulas with stepsize
  control and their application to some heat transfer problems}, Tech. rep.,
  NASA Marshall Space Flight Center, Huntsville

\bibitem[{{Fowler} \& {Hoyle}(1964)}]{1964ApJS....9..201F}
{Fowler}, W.~A., \& {Hoyle}, F. 1964, \apjs, 9, 201

\bibitem[{{Fujii} {et~al.}(2021){Fujii}, {Saitoh}, {Hirai}, \&
  {Wang}}]{2021PASJ...73.1074F}
{Fujii}, M.~S., {Saitoh}, T.~R., {Hirai}, Y., \& {Wang}, L. 2021, \pasj, 73,
  1074

\bibitem[{{Fung} {et~al.}(2014){Fung}, {Shi}, \&
  {Chiang}}]{2014ApJ...782...88F}
{Fung}, J., {Shi}, J.-M., \& {Chiang}, E. 2014, \apj, 782, 88

\bibitem[{{Gilbaum} \& {Stone}(2022)}]{2022ApJ...928..191G}
{Gilbaum}, S., \& {Stone}, N.~C. 2022, \apj, 928, 191

\bibitem[{{Ginzburg} \& {Sari}(2018)}]{2018MNRAS.479.1986G}
{Ginzburg}, S., \& {Sari}, R. 2018, \mnras, 479, 1986

\bibitem[{{Goodman} \& {Tan}(2004)}]{2004ApJ...608..108G}
{Goodman}, J., \& {Tan}, J.~C. 2004, \apj, 608, 108

\bibitem[{{Graham} {et~al.}(2017){Graham}, {Djorgovski}, {Drake}, {Stern},
  {Mahabal}, {Glikman}, {Larson}, \& {Christensen}}]{2017MNRAS.470.4112G}
{Graham}, M.~J., {Djorgovski}, S.~G., {Drake}, A.~J., {et~al.} 2017, \mnras,
  470, 4112

\bibitem[{{Grishin} {et~al.}(2021){Grishin}, {Bobrick}, {Hirai}, {Mandel}, \&
  {Perets}}]{2021MNRAS.507..156G}
{Grishin}, E., {Bobrick}, A., {Hirai}, R., {Mandel}, I., \& {Perets}, H.~B.
  2021, \mnras, 507, 156

\bibitem[{{Habibi} {et~al.}(2017){Habibi}, {Gillessen}, {Martins},
  {Eisenhauer}, {Plewa}, {Pfuhl}, {George}, {Dexter}, {Waisberg}, {Ott}, {von
  Fellenberg}, {Baub{\"o}ck}, {Jimenez-Rosales}, \& {Genzel}}]{Habibi:2017}
{Habibi}, M., {Gillessen}, S., {Martins}, F., {et~al.} 2017, \apj, 847, 120

\bibitem[{Hill(1878)}]{hill1878researches}
Hill, G.~W. 1878, American journal of Mathematics, 1, 5

\bibitem[{{Ho}(2008)}]{2008ARA&A..46..475H}
{Ho}, L.~C. 2008, \araa, 46, 475

\bibitem[{{Hogg} \& {Reynolds}(2016)}]{2016ApJ...826...40H}
{Hogg}, J.~D., \& {Reynolds}, C.~S. 2016, \apj, 826, 40

\bibitem[{{Hopkins} {et~al.}(2024){Hopkins}, {Grudic}, {Kremer}, {Offner},
  {Guszejnov}, \& {Rosen}}]{2024arXiv240408046H}
{Hopkins}, P.~F., {Grudic}, M.~Y., {Kremer}, K., {et~al.} 2024, arXiv e-prints,
  arXiv:2404.08046

\bibitem[{{Hoyle} \& {Lyttleton}(1939)}]{1939PCPS...35..405H}
{Hoyle}, F., \& {Lyttleton}, R.~A. 1939, Proceedings of the Cambridge
  Philosophical Society, 35, 405

\bibitem[{{Hunter}(2007)}]{4160265}
{Hunter}, J.~D. 2007, Computing in Science Engineering, 9, 90

\bibitem[{{Jermyn} {et~al.}(2021){Jermyn}, {Dittmann}, {Cantiello}, \&
  {Perna}}]{2021ApJ...914..105J}
{Jermyn}, A.~S., {Dittmann}, A.~J., {Cantiello}, M., \& {Perna}, R. 2021, \apj,
  914, 105

\bibitem[{{Jermyn} {et~al.}(2022){Jermyn}, {Dittmann}, {McKernan}, {Ford}, \&
  {Cantiello}}]{2022ApJ...929..133J}
{Jermyn}, A.~S., {Dittmann}, A.~J., {McKernan}, B., {Ford}, K.~E.~S., \&
  {Cantiello}, M. 2022, \apj, 929, 133

\bibitem[{{Kanagawa} {et~al.}(2015){Kanagawa}, {Tanaka}, {Muto}, {Tanigawa}, \&
  {Takeuchi}}]{2015MNRAS.448..994K}
{Kanagawa}, K.~D., {Tanaka}, H., {Muto}, T., {Tanigawa}, T., \& {Takeuchi}, T.
  2015, \mnras, 448, 994

\bibitem[{{King} {et~al.}(2007){King}, {Pringle}, \&
  {Livio}}]{2007MNRAS.376.1740K}
{King}, A.~R., {Pringle}, J.~E., \& {Livio}, M. 2007, \mnras, 376, 1740

\bibitem[{{Kolykhalov} \& {Syunyaev}(1980)}]{1980SvAL....6..357K}
{Kolykhalov}, P.~I., \& {Syunyaev}, R.~A. 1980, Soviet Astronomy Letters, 6,
  357

\bibitem[{Kutta(1901)}]{Kutta}
Kutta, W. 1901, Zeit. Math. Phys., 46, 435

\bibitem[{{Levin} \& {Beloborodov}(2003)}]{Levin:2003}
{Levin}, Y., \& {Beloborodov}, A.~M. 2003, \apjl, 590, L33

\bibitem[{{Maeder}(1987)}]{1987A&A...178..159M}
{Maeder}, A. 1987, \aap, 178, 159

\bibitem[{{McKernan} {et~al.}(2012){McKernan}, {Ford}, {Lyra}, \&
  {Perets}}]{2012MNRAS.425..460M}
{McKernan}, B., {Ford}, K.~E.~S., {Lyra}, W., \& {Perets}, H.~B. 2012, \mnras,
  425, 460

\bibitem[{{McKernan} {et~al.}(2020){McKernan}, {Ford}, \&
  {O'Shaughnessy}}]{2020MNRAS.498.4088M}
{McKernan}, B., {Ford}, K.~E.~S., \& {O'Shaughnessy}, R. 2020, \mnras, 498,
  4088

\bibitem[{{Owocki} {et~al.}(2004){Owocki}, {Gayley}, \&
  {Shaviv}}]{2004ApJ...616..525O}
{Owocki}, S.~P., {Gayley}, K.~G., \& {Shaviv}, N.~J. 2004, \apj, 616, 525

\bibitem[{{Paumard} {et~al.}(2006){Paumard}, {Genzel}, {Martins}, {Nayakshin},
  {Beloborodov}, {Levin}, {Trippe}, {Eisenhauer}, {Ott}, {Gillessen}, {Abuter},
  {Cuadra}, {Alexander}, \& {Sternberg}}]{Paumard:2006}
{Paumard}, T., {Genzel}, R., {Martins}, F., {et~al.} 2006, \apj, 643, 1011

\bibitem[{{Paxton} {et~al.}(2011){Paxton}, {Bildsten}, {Dotter}, {Herwig},
  {Lesaffre}, \& {Timmes}}]{2011ApJS..192....3P}
{Paxton}, B., {Bildsten}, L., {Dotter}, A., {et~al.} 2011, \apjs, 192, 3

\bibitem[{{Perna} {et~al.}(2021){Perna}, {Lazzati}, \&
  {Cantiello}}]{2021ApJ...906L...7P}
{Perna}, R., {Lazzati}, D., \& {Cantiello}, M. 2021, \apjl, 906, L7

\bibitem[{{Portegies Zwart} \& {van den Heuvel}(2016)}]{2016MNRAS.456.3401P}
{Portegies Zwart}, S.~F., \& {van den Heuvel}, E.~P.~J. 2016, \mnras, 456, 3401

\bibitem[{{Rosenthal} {et~al.}(2020){Rosenthal}, {Chiang}, {Ginzburg}, \&
  {Murray-Clay}}]{2020MNRAS.498.2054R}
{Rosenthal}, M.~M., {Chiang}, E.~I., {Ginzburg}, S., \& {Murray-Clay}, R.~A.
  2020, \mnras, 498, 2054

\bibitem[{{Sakurai} {et~al.}(2015){Sakurai}, {Hosokawa}, {Yoshida}, \&
  {Yorke}}]{2015MNRAS.452..755S}
{Sakurai}, Y., {Hosokawa}, T., {Yoshida}, N., \& {Yorke}, H.~W. 2015, \mnras,
  452, 755

\bibitem[{{Salpeter}(1964)}]{1964ApJ...140..796S}
{Salpeter}, E.~E. 1964, \apj, 140, 796

\bibitem[{{Shima} {et~al.}(1985){Shima}, {Matsuda}, {Takeda}, \&
  {Sawada}}]{1985MNRAS.217..367S}
{Shima}, E., {Matsuda}, T., {Takeda}, H., \& {Sawada}, K. 1985, \mnras, 217,
  367

\bibitem[{{Sirko} \& {Goodman}(2003)}]{2003MNRAS.341..501S}
{Sirko}, E., \& {Goodman}, J. 2003, \mnras, 341, 501

\bibitem[{{Smith}(2014)}]{2014ARA&A..52..487S}
{Smith}, N. 2014, \araa, 52, 487

\bibitem[{{Stone} {et~al.}(2017){Stone}, {Metzger}, \&
  {Haiman}}]{2017MNRAS.464..946S}
{Stone}, N.~C., {Metzger}, B.~D., \& {Haiman}, Z. 2017, \mnras, 464, 946

\bibitem[{{Syer} {et~al.}(1991){Syer}, {Clarke}, \&
  {Rees}}]{1991MNRAS.250..505S}
{Syer}, D., {Clarke}, C.~J., \& {Rees}, M.~J. 1991, \mnras, 250, 505

\bibitem[{{Tagawa} {et~al.}(2020){Tagawa}, {Haiman}, \&
  {Kocsis}}]{2020ApJ...898...25T}
{Tagawa}, H., {Haiman}, Z., \& {Kocsis}, B. 2020, \apj, 898, 25

\bibitem[{{Thompson} {et~al.}(2005){Thompson}, {Quataert}, \&
  {Murray}}]{2005ApJ...630..167T}
{Thompson}, T.~A., {Quataert}, E., \& {Murray}, N. 2005, \apj, 630, 167

\bibitem[{{Toyouchi} {et~al.}(2022){Toyouchi}, {Inayoshi}, {Ishigaki}, \&
  {Tominaga}}]{2022MNRAS.512.2573T}
{Toyouchi}, D., {Inayoshi}, K., {Ishigaki}, M.~N., \& {Tominaga}, N. 2022,
  \mnras, 512, 2573

\bibitem[{{van der Helm} {et~al.}(2016){van der Helm}, {Portegies Zwart}, \&
  {Pols}}]{2016MNRAS.455..462V}
{van der Helm}, E., {Portegies Zwart}, S., \& {Pols}, O. 2016, \mnras, 455, 462

\bibitem[{{van der Walt} {et~al.}(2011){van der Walt}, {Colbert}, \&
  {Varoquaux}}]{5725236}
{van der Walt}, S., {Colbert}, S.~C., \& {Varoquaux}, G. 2011, Computing in
  Science Engineering, 13, 22

\bibitem[{Virtanen {et~al.}(2020)Virtanen, Gommers, Oliphant, Haberland, Reddy,
  Cournapeau, Burovski, Peterson, Weckesser, Bright, {van der Walt}, Brett,
  Wilson, Millman, Mayorov, Nelson, Jones, Kern, Larson, Carey, Polat, Feng,
  Moore, {VanderPlas}, Laxalde, Perktold, Cimrman, Henriksen, Quintero, Harris,
  Archibald, Ribeiro, Pedregosa, {van Mulbregt}, \& {SciPy 1.0
  Contributors}}]{2020SciPy-NMeth}
Virtanen, P., Gommers, R., Oliphant, T.~E., {et~al.} 2020, Nature Methods, 17,
  261

\bibitem[{{Xu} {et~al.}(2018){Xu}, {Bian}, {Shen}, {Zuo}, {Fan}, \&
  {Zhu}}]{2018MNRAS.480..345X}
{Xu}, F., {Bian}, F., {Shen}, Y., {et~al.} 2018, \mnras, 480, 345

\bibitem[{{Yoon} \& {Langer}(2005)}]{2005A&A...443..643Y}
{Yoon}, S.~C., \& {Langer}, N. 2005, \aap, 443, 643

\bibitem[{{Zhou} {et~al.}(2024){Zhou}, {Sun}, {Liu}, {Wang}, {Wang}, \&
  {Xue}}]{2024ApJ...966L...9Z}
{Zhou}, S., {Sun}, M., {Liu}, T., {et~al.} 2024, \apjl, 966, L9

\end{thebibliography}

\end{document}